\documentclass[12pt,preprint]{emulateapj}

\usepackage{amsmath}
\usepackage{bm}

\slugcomment{\today}

\shorttitle{Protoneutron star cooling and nuclear matter}
\shortauthors{Nakazato and Suzuki}

\begin{document}

\title{Cooling timescale for protoneutron stars and properties of nuclear matter: Effective mass and symmetry energy at high densities}

\author{Ken'ichiro Nakazato\altaffilmark{1}, and Hideyuki Suzuki\altaffilmark{2}}

\email{nakazato@artsci.kyushu-u.ac.jp}

\altaffiltext{1}{Faculty of Arts \& Science, Kyushu University, 744 Motooka, Nishi-ku, Fukuoka 819-0395, Japan}
\altaffiltext{2}{Faculty of Science \& Technology, Tokyo University of Science, 2641 Yamazaki, Noda, Chiba 278-8510, Japan}

\begin{abstract}
The cooling process of a protoneutron star is investigated with focus on its sensitivity to properties of hot and dense matter. An equation of state, which includes the nucleon effective mass and nuclear symmetry energy at twice the saturation density as control parameters, is constructed for systematic studies. The numerical code utilized in this study follows a quasi-static evolution of a protoneutron star solving the general-relativistic stellar structure with neutrino diffusion. The cooling timescale evaluated from the neutrino light curve is found to be longer for the models with larger effective masses and smaller symmetry energies at high densities. The present results are compared with those for other equations of state and it is found that they are consistent in terms of their dependences on the effective mass and neutron star radius.
\end{abstract}


\section{Introduction}\label{sec:intro}
A core-collapse supernova explosion, which is the fate of massive stars with an initial mass of $\gtrsim$10$M_\odot$, is one of the most complicated but, therefore, interesting phenomena in the Universe because it involves many physical ingredients. Inside the supernova core, the matter density exceeds the nuclear saturation density and the temperature is on the order of 10~MeV and above. After the explosion, a protoneutron star (PNS), which is a newly formed hot compact object, remains. Meanwhile, numerous neutrinos of all flavors are emitted from core-collapse supernovae. The evolutions of the supernova core and PNS are governed by the properties of the hot and dense nuclear matter through an equation of state (EOS), and the thermodynamic condition is affected by the reactions and diffusion of neutrinos. Thus, the nuclear EOS and neutrinos play key roles in core-collapse supernovae. Nevertheless, their details are still not well understood and many theoretical studies employing numerical simulations have been devoted to this field \citep[e.g.,][]{kotake12,burrows13,self13a,foglizzo15,janka16,rr17}.

The neutrino emission from core-collapse supernovae can be divided into three phases: (i) the neutronization burst, (ii) the accretion phase and (iii) the cooling phase. Since the core of a massive star becomes gravitationally unstable at the end of its evolution, the core starts to collapse and the collapse is bounced by the nuclear repulsion force. A shock wave is launched by the bounced inner core and dissociates nuclei in the outer core producing a large amount of $\nu_e$ via electron capture. This is called the neutronization burst. Subsequently, the outer core matter accretes onto the inner core releasing the gravitational potential energy as neutrinos, which is the accretion phase that lasts until the core explosion. The duration of this phase is thought to be less than 1~s \citep[][]{totani98,fischer10,hude10,suwa14}. After the shock wave successfully propagates into the stellar envelope, neutrinos carry away thermal energy from the remnant PNS. This is called the cooling phase, and we focus on this phase in this paper.

The observation of supernova neutrinos provides us with a unique opportunity to probe inside the supernova core and PNS. As is well known, neutrinos emitted from SN1987A were detected by the neutrino detectors Kamiokande II \citep[][]{hirata87}, IMB \citep[][]{bionta87} and Baksan \citep[][]{alex88}. The duration of neutrino detection was $\sim$10~s; therefore, the cooling timescale of a PNS should be comparable to or longer than 10~s \citep[][]{spergel87,suzuki88,loredo02,pagl09}. In the case of SN1987A, the progenitor resided in the Large Magellanic Cloud, which is at a distance of 50~kpc, and about 20 events were detected in total. In contrast, SuperKamiokande, which is a currently operating water Cerenkov neutrino detector, will detect about 10,000 events for the next Galactic core-collapse supernova. Furthermore, the duration of neutrino detection will be longer than 10~s \citep[][]{suwa19}.

So as to evaluate the long-term variation of supernova neutrino signals, numerical simulations of the PNS cooling have been employed \citep[e.g.,][]{burr86,suzuki94,pons99,robe12b}. Previous studies have confirmed that neutrino emission continues for at least 10~s, which can account for the observational data of SN1987A. However, it has also been found that the behavior of neutrino emission in later epochs depends on the underlying nuclear EOS \citep[][]{sumi95,pons99,robe12,came17,self18}. Although numerical simulations of PNS cooling require an EOS including finite temperatures, currently available models based on nuclear theories are limited \citep[e.g.,][]{oertel17}. In this study, we investigate the EOS dependence of PNS cooling by constructing a new series of phenomenological EOSs. In particular, we focus on the cooling timescale, which corresponds to the duration of emission.

Recently, properties of nuclear matter in the vicinity of the saturation density have been constrained by several terrestrial experiments \citep[e.g.,][]{shlomo06,khan12,latlim13,tews17}. In particular for symmetric nuclear matter, which is composed of the same number of protons as neutrons, the incompressibility parameter is determined with an uncertainty of $\sim$20\%. On the other hand, the behavior of neutron-rich matter at high densities remains controversial. Since neutron stars contain neutron-rich matter, the structure of neutron stars, such as the radius, is sensitive to the EOS. Therefore, observations of neutron stars contribute to resolving this issue \citep[e.g.,][]{steiner10,sotani12,LIGO18}. Neutrino detection from PNS cooling may be useful for probing not only dense but also hot nuclear matter. In particular, the thermal properties of nuclear matter are determined by the nucleon effective mass \citep[][]{const15}. In this study, we introduce the effective mass into our model so as to examine its impact on PNS cooling.

The purpose of this study is to investigate the EOS dependence of PNS cooling. In \S~\ref{sec:eos}, we describe our EOS model used in this study. For the convenience of theoretical analysis and physics considerations, we adopt the phenomenological expansion form with empirical parameters. While this approach is simple, it enable us to obtain results independent of the microscopic details of nuclear interactions. In \S~\ref{sec:pns}, we introduce the numerical methods and PNS models used in our cooling simulations. The main results are shown in \S~\ref{sec:rad}. So as to evaluate the cooling timescale, we employ an $e$-folding time of the neutrino luminosity and its maximum value. Furthermore, we compare the cooling timescale of our EOS models with those of other supernova EOSs in relation to the neutron star radius as well as the effective mass. Finally, \S~\ref{sec:conc} is devoted to our conclusions.

\section{Construction of EOS}\label{sec:eos}

We construct a series of phenomenological EOSs to apply the numerical simulations of PNS cooling. In this study, the EOS for zero-temperature matter and the effect of finite temperatures are considered separately. In \S~\ref{sec:0eos}, the zero-temperature EOS, which mimics the properties of nuclear matter at high densities, is described. We introduce the method to deal with the effect of finite temperatures in \S~\ref{sec:teos}. Below the saturation density, nuclear matter undergoes a phase transition from a uniform phase to an inhomogeneous phase. In \S~\ref{sec:ieos}, our model for the low-density region where an inhomogeneous phase appears is illustrated. The mass--radius relations of cold neutron stars based on our EOS are also shown in \S~\ref{sec:nseos} so as to examine their consistency with the results of a recent analysis of the gravitational wave from a neutron star merger \citep[][]{LIGO18}.

\subsection{Zero temperature}\label{sec:0eos}
In describing the energy of uniform dense matter at zero temperature as a function of baryon number density $n_b$ and proton fraction $Y_p$, we employ the following simple expression for the energy per baryon:
\begin{equation}
w(n_b,Y_p) = w_0 + \frac{K_0}{18n_0^2} (n_b-n_0)^2 + S(n_b)\,(1-2Y_p)^2,
\label{eq:symeos}
\end{equation}
where $n_0$, $w_0$ and $K_0$ are, respectively, the saturation density, saturation energy and incompressibility of symmetric nuclear matter, which is matter with $Y_p=0.5$. Throughout this paper, we set $n_0=0.16$~fm$^{-3}$, $w_0=-16$~MeV and $K_0=245$~MeV. The symmetry energy $S(n_b)$, which is the energy difference between symmetric nuclear matter ($Y_p=0.5$) and pure neutron matter ($Y_p=0$), is considered as a function of $n_b$ in this study. As usual, we assume that $w(n_b,Y_p)$ increases quadratically with $Y_p$ from $Y_p=0.5$ to $Y_p=0$ \citep[e.g.,][]{oyak03}.

For the symmetry energy $S(n_b)$, we again adopt a simple polynomial expression. Below the saturation density, it is written as
\begin{equation}
S(n_b) = S_0 + \frac{L}{3n_0} (n_b-n_0) + \frac{K_{\rm sym}}{18n_0^2} (n_b-n_0)^2,
\label{eq:lsymene}
\end{equation}
where $S_0$ and $L$ are the coefficients of the symmetry energy and the symmetry energy density derivative at the saturation density, respectively. Throughout this paper, we set $S_0=31$~MeV and $L=50$~MeV \citep[e.g.,][]{tews17}. The coefficient of a higher-derivative term, $K_{\rm sym}$, is difficult to measure experimentally. On the basis of the neutron matter calculations by \citet{dris16}, $-240$~${\rm MeV} < K_{\rm sym} < -70$~MeV is suggested. In this study, we choose $K_{\rm sym}=-150$~MeV.

In contrast to the case below the saturation density, we write the symmetry energy above the saturation density as
\begin{equation}
S(n_b) = S_0 + \frac{L}{3n_0} (n_b-n_0) + \frac{1}{n_0^2} \left( S_{00} - S_0 -\frac{L}{3} \right)(n_b-n_0)^2,
\label{eq:hsymene}
\end{equation}
where $S_0$ and $L$ are the same parameters as in equation~(\ref{eq:lsymene}). In this expression, $S_{00}$ is the symmetry energy at the density of $2n_0$: $S(2n_0) = S_{00}$. Hereafter, we employ $S_{00}$ as a parameter with which the symmetry energy at high densities is characterized and we investigate its impacts on PNS cooling without changing the EOS at subsaturation densities. Note that, although equations~(\ref{eq:lsymene}) and (\ref{eq:hsymene}) are mathematically equivalent, their spirits are different. Equation~(\ref{eq:lsymene}) is a second-order Taylor expansion of the density-dependent symmetry energy around $n_b=n_0$ and is intended to feasibly describe the EOS at subsaturation densities. On the other hand, equation~(\ref{eq:hsymene}) is a gross approximation for supranuclear densities by a quadratic form.

Owing to the simple expressions, we can easily calculate the energy density and pressure of baryons at zero temperature with $\varepsilon^{(0)}_b=n_bw$ and $P^{(0)}_b=n_b^2\frac{\partial}{\partial n_b}w$, respectively. Furthermore, with a neutron number density $n_n=(1-Y_p)n_b$ and proton number density $n_p=Y_pn_b$, the chemical potentials of neutrons and protons are obtained analytically using $\mu^{(0)}_n=\left(\frac{\partial}{\partial n_n}\varepsilon^{(0)}_b \right)_{n_p}$ and $\mu^{(0)}_p=\left(\frac{\partial}{\partial n_p}\varepsilon^{(0)}_b \right)_{n_n}$, respectively.

~\\

\subsection{Finite temperatures}\label{sec:teos}
So as to take into account the effect of finite temperatures, we employ the thermodynamic quantities of an ideal Fermi gas. We denote by $\varepsilon^{\rm F}_i(n_i,T;M^\ast_i)$ the energy density of ideal fermions $i$ with number density $n_i$, temperature $T$ and effective mass $M^\ast_i$. Similarly, the pressure and entropy per particle of ideal fermions $i$ are denoted by $P^{\rm F}_i(n_i,T;M^\ast_i)$ and $s^{\rm F}_i(n_i,T;M^\ast_i)$, respectively. In considering the thermal effects of neutrons $n$ and protons $p$, the Helmholtz free energy per baryon $F_b$, pressure $P_b$ and entropy per baryon $s_b$ of baryons at temperature $T$ are written as 
\begin{subequations}
\begin{gather}
\begin{split}
  F_b(n_b,Y_p,T) =& \, \frac{1}{n_b}\Bigl[\varepsilon^{(0)}_b(n_b,Y_p) \\
   & \qquad + \varepsilon^{\rm F}_n(n_n,T;M^\ast_n)-\varepsilon^{\rm F}_n(n_n,0;M^\ast_n) \\
    & \qquad + \varepsilon^{\rm F}_p(n_p,T;M^\ast_p)-\varepsilon^{\rm F}_p(n_p,0;M^\ast_p)\Bigr] \\
  & -Ts_b(n_b,Y_p,T),
\end{split}
\label{eq:hfeb}
\end{gather}
\begin{gather}
\begin{split}
  P_b(n_b,Y_p,T) =& \, P^{(0)}_b(n_b,Y_p) \\
  & +P^{\rm F}_n(n_n,T;M^\ast_n)-P^{\rm F}_n(n_n,0;M^\ast_n) \\
  & +P^{\rm F}_p(n_p,T;M^\ast_p)-P^{\rm F}_p(n_p,0;M^\ast_p),
\label{eq:preb}
\end{split}
\end{gather}
\begin{equation}
s_b(n_b,Y_p,T) = (1-Y_p)s^{\rm F}_n(n_n,T;M^\ast_n) + Y_ps^{\rm F}_p(n_p,T;M^\ast_p),
\label{eq:entb}
\end{equation}
\label{eq:eosb}
\end{subequations}
where $M^\ast_n$ and $M^\ast_p$ are the effective masses of neutrons and protons, respectively. Whereas the effective masses depend on the number density in general, for simplicity we set them to be constant in this study similarly to in \citet{ls91}. Similarly, we write the chemical potentials of neutrons and protons as
\begin{subequations}
\begin{gather}
\begin{split}
  \mu_n(n_b,Y_p,T) =& \, \mu^{(0)}_n(n_b,Y_p) \\
  & + \mu^{\rm F}_n(n_n,T;M^\ast_n) - \mu^{\rm F}_n(n_n,0;M^\ast_n),
\label{eq:cptn}
\end{split}
\end{gather}
\begin{gather}
\begin{split}
  \mu_p(n_b,Y_p,T) =& \, \mu^{(0)}_p(n_b,Y_p) \\
  & + \mu^{\rm F}_p(n_p,T;M^\ast_p) - \mu^{\rm F}_p(n_p,0;M^\ast_p),
\label{eq:cptp}
\end{split}
\end{gather}
\label{eq:cptb}
\end{subequations}
where $\mu^{\rm F}_i(n_i,T;M^\ast_i)$ is the chemical potential of ideal fermions $i$. Furthermore, for our EOS, we add the contributions of photons, electrons and positrons to satisfy the charge neutrality.

In the above expressions, the zero-temperature EOS does not depend on the choice of $M^\ast_n$ and $M^\ast_p$. In this study, we assume that the effective mass in units of rest mass is the same for neutrons and protons and we denote it by $u$: $u=M^\ast_n/M_n=M^\ast_p/M_p$ with neutron rest mass $M_n$ and proton rest mass $M_p$. In the following investigation of PNS cooling, we employ $u$ as a parameter with which the effect of finite temperatures is characterized. In this way, our EOS is advantageous for studying the dependence of the cooling timescale separating the roles of the effective mass and symmetry energy at high densities.

~\\

\subsection{Inhomogeneous matter}\label{sec:ieos}

While the EOS shown in \S~\ref{sec:0eos} and \ref{sec:teos} is for uniform nuclear matter, the phase transition to inhomogeneous matter occurs at subsaturation densities. In this study, we utilize the inhomogeneous matter EOS constructed by \citet{shen11}. For this purpose, our uniform EOS and the Shen EOS are connected by the same method as in \citet{self18}. We adopt our uniform EOS for the baryon mass density $\rho_b \ge 10^{14.3}$~g~cm$^{-3}$ and the Shen EOS for $\rho_b \le 10^{14}$~g~cm$^{-3}$. Then, by interpolating these two EOSs, we obtain thermodynamic quantities, such as the free energy, for 10$^{14}$~g~cm$^{-3} < \rho_b < 10^{14.3}$~g~cm$^{-3}$. In practice, the values in our uniform EOS and the Shen EOS with the same temperature and proton fraction are interpolated in the density direction. Incidentally, the saturation density $n_0=0.16$~fm$^{-3}$ corresponds to the baryon mass density of $10^{14.42}$~g~cm$^{-3}\approx 2.66\times 10^{14}$~g~cm$^{-3}$.

Actually, the properties of inhomogeneous matter, such as the mass number of heavy nuclei, depend on the nuclear symmetry energy \citep[e.g.,][]{oyak03,oyak07} and affect neutrino emission from a PNS. Nevertheless, \citet{self18} showed that the neutrino luminosities are insensitive to inhomogeneous matter, at least until the luminosities of individual neutrino species drop to $10^{50}$~erg~s$^{-1}$. In this study, we focus on neutrino luminosities to investigate the cooling timescale of a PNS.

\subsection{Application to cold neutron stars}\label{sec:nseos}

\begin{figure*}[t]
\plotone{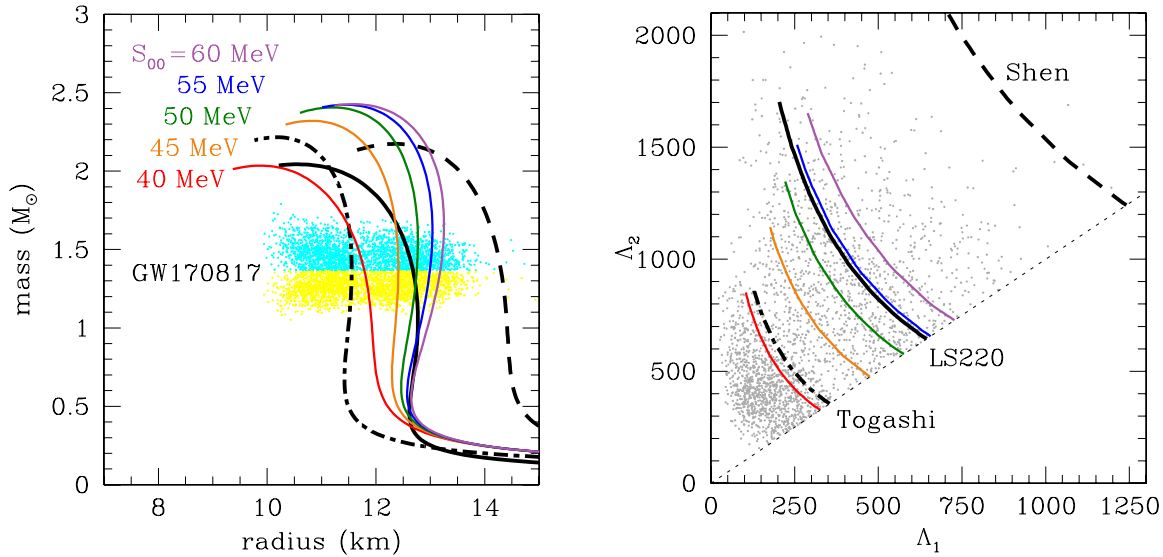}
\epsscale{1.0}
\caption{Mass--radius relation of cold neutron stars (left) and tidal deformability parameters of neutron stars in a binary with a chirp mass of $1.188M_\odot$ (right). In the right panel, $\Lambda_1$ and $\Lambda_2$ are tidal deformability parameters of the primary and secondary neutron stars, respectively. In both panels, the thin solid lines correspond, from left to right, to EOSs with $S_{00}=40$, 45, 50, 55 and 60~MeV constructed in this study. Thick solid, dashed and dot-dashed lines correspond respectively to the LS220 EOS, Shen EOS and Togashi EOS. The dots show the constraints from GW170817 taken from \citet{LIGO18}.}
\label{fig:cns}
\end{figure*}

In this subsection, we calculate the structure of neutron stars to examine the behavior of our EOS described above. Since neutron stars are well modeled with zero-temperature matter, we study the dependence of the neutron star structure solely on $S_{00}$. We assume that the neutron star matter is composed of neutrons, protons, electrons and muons satisfying charge neutrality and $\beta$-equilibrium. In some cases, pure neutron matter could be stable at very high densities since the symmetry energy becomes negative. For comparison, we also examine other EOSs constructed for supernova simulations: the Lattimer and Swesty EOS with incompressibility of 220~MeV \citep[LS220 EOS;][]{ls91}, the Shen EOS \citep[][]{shen98a,shen98b,shen11} and the Togashi EOS \citep[][]{togashi17}. While the former two EOSs are commonly used models, the Togashi EOS is a model recently constructed on the basis of variational many-body theory with the AV18 two-nucleon potential and UIX three-nucleon potential \citep[][]{kann07,kann09,togashi13}.

In Figure~\ref{fig:cns}, we show the mass--radius relation of neutron stars for our EOS. Hereafter, we examine the cases where $S_{00}=40$, 45, 50, 55 and 60~MeV for the symmetry energy at a density of $2n_0$. Then, the maximum mass of neutron stars is above $2M_\odot$ for all considered models, which is consistent with recent observations \citep[][]{demo10,antoni13,nanogra18}. If we choose a small value for $S_{00}$, the EOS becomes soft and the maximum mass becomes low. While our EOS becomes acausal at extremely high densities due to its nonrelativistic treatment, causality is satisfied at densities below the central density of neutron stars with mass less than $2M_\odot$. Incidentally, pure neutron matter appears in the central region for the model with $S_{00}=40$~MeV and $m \ge 1.72M_\odot$, where $m$ is the gravitational mass of the neutron star. Note that we investigate the cooling of a PNS with $m \sim 1.33M_\odot$ in this paper.

Since our EOS models share the saturation properties characterized by $n_0$, $w_0$, $K_0$, $S_0$, $L$ and $K_{\rm sym}$, the mass--radius relation is insensitive to EOS for low-mass neutron stars. Nevertheless, the radius of a neutron star with $m \gtrsim 1.2M_\odot$ certainly depends on $S_{00}$. The model with a large $S_{00}$ has neutron stars with a large radius because the EOS is stiff. Recently, the detection of gravitational waves from the binary neutron star merger GW170817 has given a new constraint on the neutron star structure \citep[][]{LIGO17a,LIGO17b,LIGO18}. The mass--radius relation of our EOS is consistent with the observation of gravitational waves for all considered models (Figure~\ref{fig:cns}). We also calculate the tidal deformability parameter of neutron stars. Since the chirp mass $m_c$ of GW170817 is well determined, we examine the combination of tidal deformability parameters for binary neutron stars with $m_c = 1.188M_\odot$ and we can see that our EOS is again consistent with the observation of gravitational waves. Incidentally, an equal-mass binary with a mass of $m=1.365M_\odot$ for each neutron star has a chirp mass of $m_c = 1.188M_\odot$. For the other EOSs, while the LS220 EOS and Togashi EOS are consistent with the observation of gravitational waves, the radius and tidal deformability parameter of the Shen EOS are quite large owing to its stiffness. 

\section{Setup of PNS cooling}\label{sec:pns}

In this section, we introduce the setup of our simulations for PNS cooling. In \S~\ref{sec:nume}, we describe the numerical method used in this study. The initial conditions and EOSs adopted here are shown in \S~\ref{sec:simdl}.

~\\

\subsection{Numerical method}\label{sec:nume}

Using our numerical code for simulations of PNS cooling \citep[][]{suzuki94}, we compute the quasi-static evolutions of a spherically symmetric PNS with neither additional mass accretion \citep[e.g.,][]{burrows88} nor convection \citep[e.g.,][]{robe12b}. The basic prescription is similar to ordinary stellar evolution codes using a Henyey-type method. The general-relativistic structure of the PNS in hydrostatic equilibrium (radius, density and metric function as functions of enclosed baryon mass) is calculated for each time step by solving the Oppenheimer--Volkoff equation and related equations. The time evolution is driven by the exchange of energy and lepton number between the matter and neutrinos due to various neutrino interactions. The resultant change in pressure causes the time evolution of the PNS structure. Simultaneously, the energy and lepton number are transported by neutrinos. For the neutrino transfer, we adopt the multigroup flux-limited diffusion scheme. The neutrino number density per energy is computed at each radial grid and time step. The energy dependence of neutrino transport coefficients is handled with the multigroup scheme. A flux limiter \citep[][]{bw82,mayle87} is introduced to manage the transparent regime with a diffusion approximation. This is necessary because the diffusion flux, which is originally proportional to the mean free path and the gradient of the neutrino number density, becomes unphysical in a transparent region having a very large mean free path. In addition, we also include general-relativistic effects such as time dilation and gravitational redshift of the neutrino energy. The three neutrino species, $\nu_e$, $\bar{\nu}_e$, and $\nu_x$, are treated separately, $\nu_x$ representing the average of $\nu_{\mu}$, $\bar{\nu}_{\mu}$, $\nu_{\tau}$, and $\bar{\nu}_{\tau}$.

The following neutrino interactions are considered in our code:
electron capture on a proton (${\rm p}\,{\rm e}^- \leftrightarrow {\rm  n}\,\nu_e$),
positron capture on a neutron (${\rm n}\,{\rm e}^+ \leftrightarrow {\rm p}\,\bar{\nu}_e$),
electron capture on a nucleus (${\rm A}\,{\rm e}^- \leftrightarrow {\rm A'}\,\nu_e$),
electron--positron pair annihilation (${\rm e}^-\,{\rm e}^+ \leftrightarrow \nu\,\bar{\nu}$),
plasmon decay ($\gamma^* \leftrightarrow \nu\,\bar{\nu}$),
nucleon bremsstrahlung (${\rm N}\,{\rm N}' \leftrightarrow {\rm N}\,{\rm N}'\,\nu\,\bar{\nu}$),
scattering off an electron/positron (${\rm e}^{\pm}\,\nu \leftrightarrow {\rm e}^{\pm}\,\nu$),
scattering off a nucleon (${\rm N}\,\nu \leftrightarrow {\rm N}\,\nu$) and
coherent scattering off a heavy nucleus (${\rm A}\,\nu \leftrightarrow {\rm A}\,\nu$),
where $\nu$ represents all species of neutrinos, A is a representative heavy nucleus, and N is either a proton or a neutron.

Most of the interaction rates are taken from
\citet{bruenn85}. Concretely, the neutrino mean free path and the
neutrino emissivity of charged current interactions with nucleons are
treated as functions of the neutrino energy, the matter temperature, the
density and the electron fraction. Pauli blocking for electrons and
nucleons in final states is incorporated using their chemical
potentials. Electron captures on representative heavy nuclei are taken
into account only for nuclei with the proton number greater than 20 and
with the neutron number less than 40 as a zero-order shell model in
\citet{bruenn85}. As for the isoenergetic scattering off nucleons, the
mean free path is proportional to an integration of the product of the
nucleon distribution function $f_{\rm N}(p_{\rm N})$ and the blocking factor
$(1-f_{\rm N}(p_{\rm N}))$. Note that, in the case of ideal non-interacting nucleon
gas, the nucleon number density $n_{\rm N}$ is related to the nucleon chemical
potential $\mu_{\rm N}$ as $n_{\rm N} = \frac{2}{(2\pi)^3}\int f_{\rm N}^0(p_{\rm N})d^3p_{\rm N}$
where $f_N^0\equiv \frac{1}{\exp(\sqrt{m_{\rm N}^2+p_{\rm N}^2}-\mu_{\rm
 N})/k_{\rm B}T)+1}$ and
$k_{\rm B}$ is the Boltzmann constant. Therefore we evaluate the
integration in the case of interacting nucleons using the bare nucleon
mass $m_{\rm N}$ and the chmical potential from the adopted EOS as $n_{\rm N}
\int f_{\rm N}^0(1-f_{\rm N}^0)d^3p_{\rm N}/\int f_{\rm N}^0d^3p_{\rm
  N}$. The mean free path for the coherent scattering off heavy nuclei
is calculated using the number density, the proton number and the
neutron number of the representative heavy nuclei as in
\citet{bruenn85}. Interaction rates for electron scattering require
integrations of products of electron distribution function, blocking
factor and reaction kernel by electron energy. To do this efficiently,
we made use of prepared numerical tables of the integrals for
dimensionless incident neutrino energy, scattered neutrino energy and
electron chemical potential in units of the temperature. Electron--positron pair annihilation is treated in the similar way using a numerical table of integrals for dimensionless neutrino energy, anti--neutrino energy and electron chemical potential.

In addition to the reactions in \citet{bruenn85}, we include neutrino
pair processes via nucleon bremsstrahlung \citep[][]{suzuki93} and
plasmon decay \citep[][]{kohyama86} in a simple manner. As for the less
important plasmon decay, the rate of the electon--positron pair process
is just enhanced by a ratio of total energy loss rates due to the two
processes. Reaction kernel for the nucleon bremsstrahlung is evaluated
using the one pion exchange model for nucleon interactions, and the low
energy limit for neutrinos and the degenerate/non-degenerate limit for
nucleons are assumed. The multiple scattering suppression effects on the
nucleon bremsstrahlung process are also taken into account as in
\citet{rs91}. Other many-body effects on neutrino opacity such as the
effective mass dependence in dense medium \citep[][]{benlov17,came17} are not included in the present study. In particular, while the dispersion relations used to construct the EOS are adopted for the calculations of neutrino opacities in \citet{pons99} and \citet{came17}, we do not assume a specific dispersion relation in the construction of our EOS. Thus, the reconstruction of the dispersion relations which are consistent with our EOS is non-trivial and we defer the investigation including the consistency between the dispersion relations used for the neutrino opacities and the EOS. 

\subsection{Simulation models}\label{sec:simdl}

\begin{figure}
\plotone{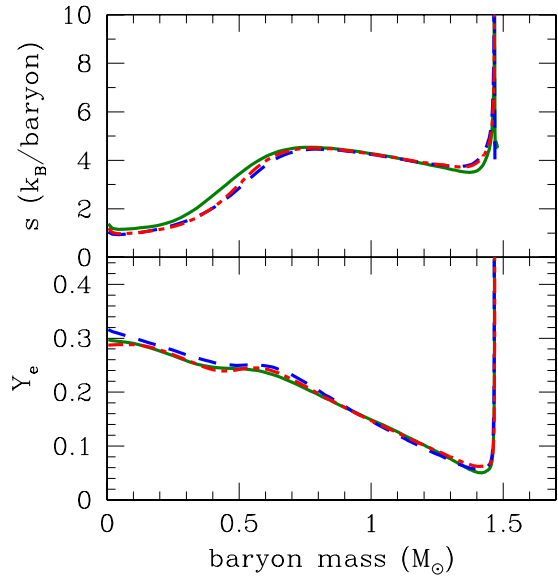}
\epsscale{1.0}
\caption{Initial profiles of entropy (upper plots) and electron fraction (lower plots) for the simulations of PNS cooling. Dot-dashed (red) lines are for the models with our EOS and the T+S EOS. Solid (green) and dashed (blue) lines correspond to the LS220 EOS and Shen EOS, respectively.}
\label{fig:init}
\end{figure}

For use as the initial conditions of our simulations of PNS cooling, we adopt the numerical results of stellar collapse as in \citet{self13a,self18}. A core collapse of the $15M_\odot$ progenitor model \citep[][]{woosley95} is followed by the numerical code of general-relativistic neutrino-radiation hydrodynamics \citep[][]{sumi05} with the Togashi EOS. Then, for the initial condition of our PNS cooling, we adopt the entropy and electron-fraction profiles at 0.3~s after the bounce, when the shock wave is stalled at the baryon mass coordinate of $m_b=1.47M_\odot$, as shown in Figure~\ref{fig:init}. Accordingly, the cooling of a PNS with a baryon mass of $1.47M_\odot$ is computed using our EOS described in \S~\ref{sec:eos}. Hereafter, we choose $u=0.5$, 0.75 and 1 for the effective mass in units of rest mass and investigate 15 models in total combining the choice of $S_{00}$ given in \S~\ref{sec:nseos}. In the following, we compute PNS cooling until the luminosity of $\bar\nu_e$ drops to $5\times10^{48}$~erg~s$^{-1}$.

\begin{figure*}
\plotone{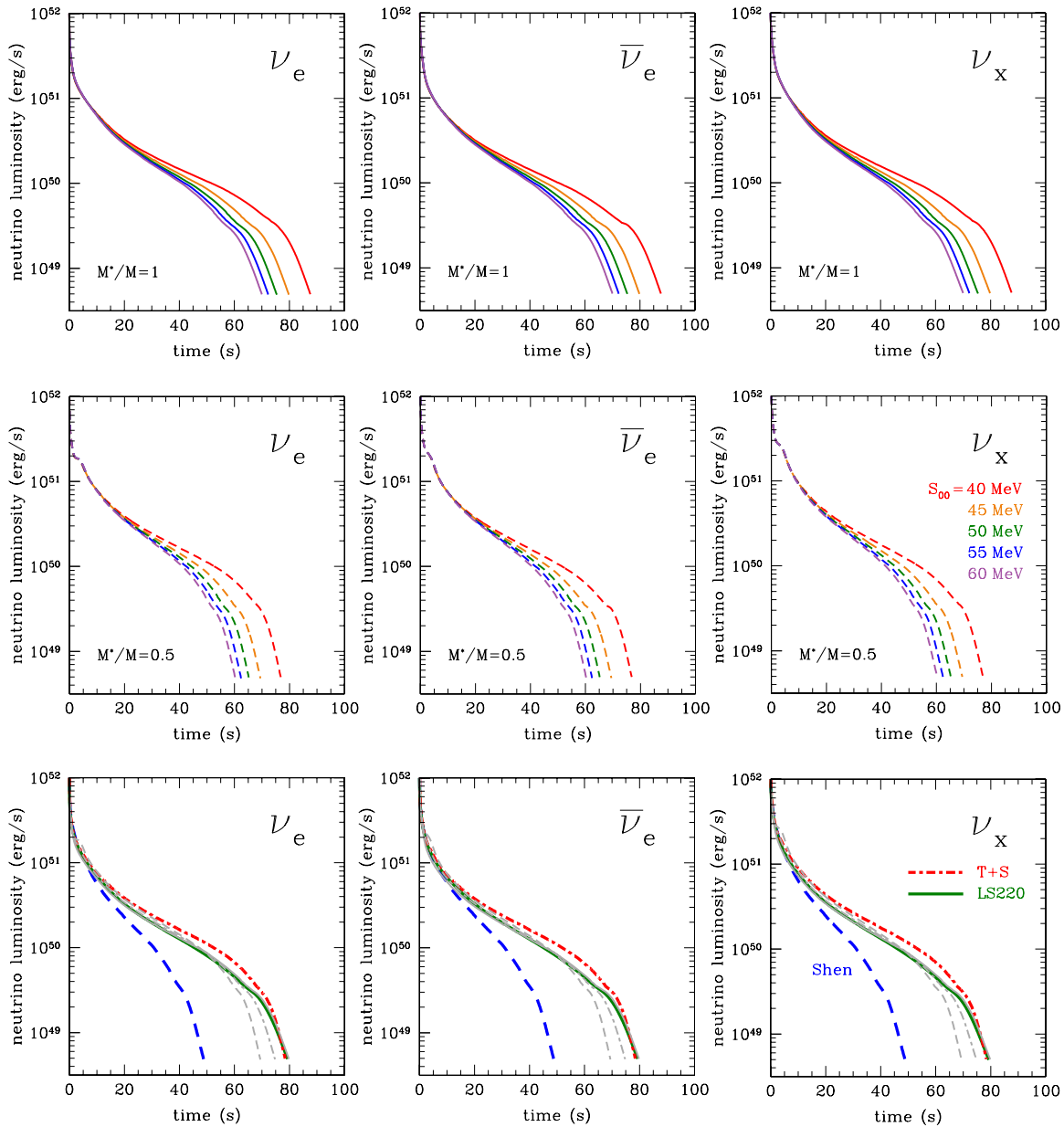}
\epsscale{1.0}
\caption{Luminosities of the emitted $\nu_e$ (left), $\bar\nu_e$ (center) and $\nu_x$ (right) as a function of time, where $\nu_x=\nu_\mu$, $\nu_\tau$, $\bar\nu_\mu$, $\bar\nu_\tau$. The upper and middle panels are for our EOS models with effective masses of $u=1$ (solid) and 0.5 (dashed), respectively, and the lines correspond, from top to bottom, to $S_{00}=40$, 45, 50, 55 and 60~MeV in these panels. In the lower panels, thick solid (green), thick dashed (blue) and thick dot-dashed (red) lines are for the LS220 EOS, Shen EOS, and T+S EOS models, respectively, whereas thin (gray) lines are for our EOS models with $S_{00}=45$~MeV and effective masses of $u=1$ (solid), 0.75 (dot-dashed), and 0.5 (dashed).}
\label{fig:nulc}
\end{figure*}

Here, we also consider PNS cooling using other supernova EOSs: the LS220 EOS, Shen EOS, and Togashi EOS. So as to prepare their initial conditions individually, core-collapse simulations using these EOSs are carried out until the central PNS grows to a baryon mass of $1.47M_\odot$. However, we find that the EOS dependence of the entropy and electron-fraction profiles is negligible (Figure~\ref{fig:init}). Incidentally, the cooling results obtained using the Shen EOS and Togashi EOS have already been reported by \citet{self18}. Furthermore, PNS cooling has also been studied using the hybrid EOS (T+S EOS), which is the same as the Togashi EOS at high densities and the Shen EOS at low densities including the inhomogeneous matter phase. Since the neutrino light curve of the Togashi EOS model suffers from numerical fluctuations, we use the T+S EOS model for comparison with the result of our EOS in this paper. Note that the time $t$ is measured from the onset of the PNS cooling simulation in this paper whereas the time after the bounce is used by \citet{self18}.

The fate of the PNS models considered in this study is a cold neutron star with a baryon mass of $1.47M_\odot$. Nevertheless, the final gravitational mass $m$ varies among the models because the relation between the baryon mass and gravitational mass depends on the EOS used. A neutron star with a baryon mass of $1.47M_\odot$ has $m=1.334M_\odot$, $1.347M_\odot$ and $1.323M_\odot$ for the LS220 EOS, Shen EOS, and T+S EOS, respectively. For our EOS models with $S_{00}=40$--60~MeV, the gravitational mass is $m=1.327$--$1.336M_\odot$.

\section{Results and discussion}\label{sec:rad}

\begin{figure*}
\plotone{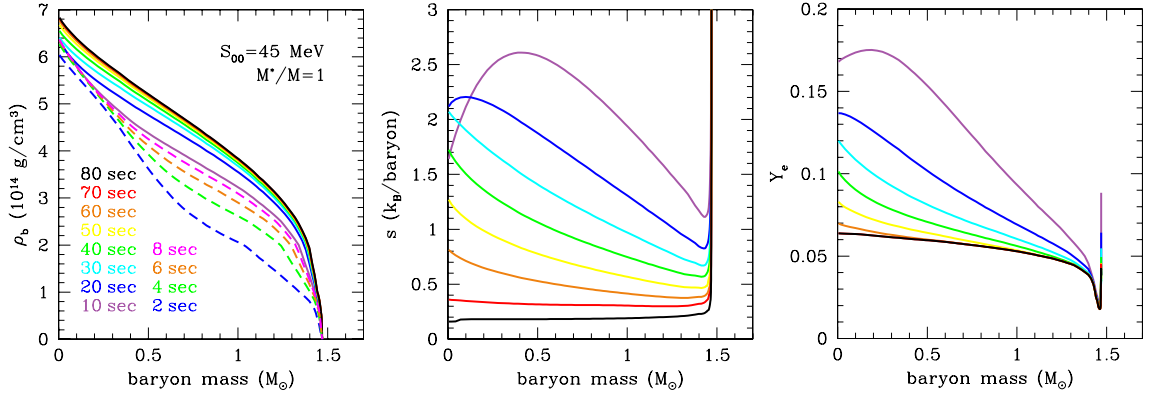}
\epsscale{1.0}
\caption{Snapshots of the PNS cooling for our EOS model with $S_{00}=45$~MeV and $u=1$. In the left panel, the density profiles are shown and the solid lines correspond, from top to bottom, to $t=80$, 70, $\ldots$, 10~s at intervals of 10~s, whereas the dashed lines correspond, from top to bottom, to $t=8$, 6, 4 and 2~s. The entropy and electron-fraction profiles are shown respectively in the central and right panels, where the lines correspond, from top to bottom, to $t=10$, 20, $\ldots$, 80~s at intervals of 10~s.}
\label{fig:snap}
\end{figure*}

\begin{figure*}
\plotone{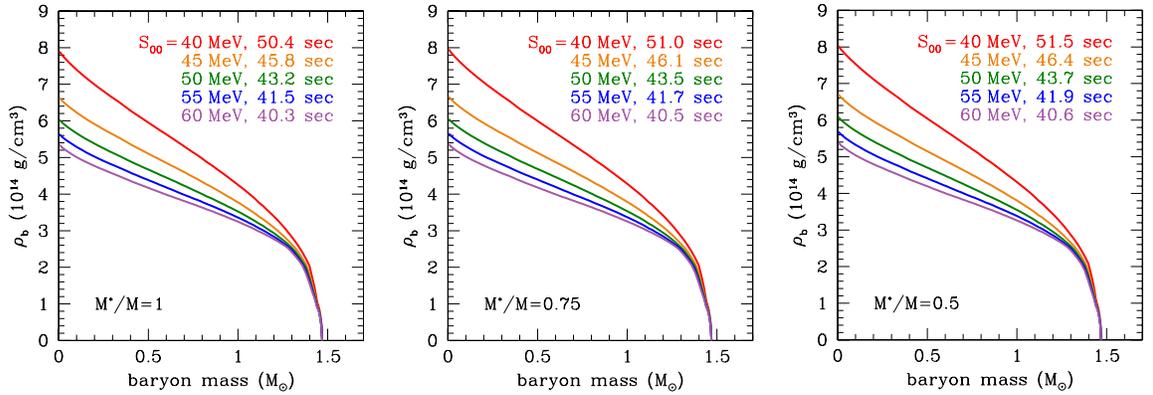}
\epsscale{1.0}
\caption{Density profiles of our EOS models with effective masses of $u=1$ (left), 0.75 (center) and 0.5 (right) at the time when the luminosity of $\bar\nu_e$ is $10^{50}$~erg~s$^{-1}$. In all panels, the lines correspond, from top to bottom, to $S_{00}=40$, 45, 50, 55 and 60~MeV.}
\label{fig:dens}
\end{figure*}

\begin{figure*}
\plotone{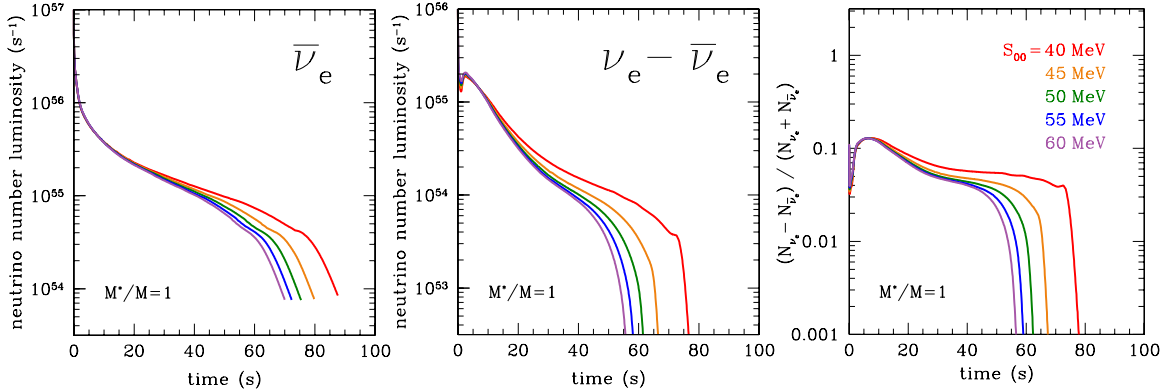}
\epsscale{1.0}
\caption{Number luminosity of $\bar\nu_e$ (left), the difference between the number luminosities of $\nu_e$ and $\bar\nu_e$ (center) and the difference between the number luminosities of $\nu_e$ and $\bar\nu_e$ divided by their sum (right) for the models with the effective mass of $u=1$. In all panels, the lines correspond, from top to bottom, to $S_{00}=40$, 45, 50, 55 and 60~MeV.}
\label{fig:nulcd}
\end{figure*}

In Figure~\ref{fig:nulc}, we show the neutrino light curves, which are the neutrino luminosities as a function of time, for the models investigated in this paper. We can see that the $\bar \nu_e$ luminosity of each model drops to $5\times10^{48}$~erg~s$^{-1}$ at most within 100~s and the luminosity is insensitive to the neutrino species especially in the late phase. Here, we use the model with $S_{00}=45$~MeV and $u=1$ as a reference model and we show its profiles of baryon mass density $\rho_b$, entropy per baryon $s$ and electron fraction $Y_e$ in Figure~\ref{fig:snap}. Until about 20~s, the PNS shrinks substantially as shown by the time dependence of the $\rho_b$ profile. In fact, the PNS radii of this model are 16.8~km, 13.8~km and 13.1~km at $t=2$~s, 10~s and 20~s, respectively. In this phase, the neutrino luminosity decreases steeply with the surface area of the PNS as well as the surface temperature. From $\sim$20~s to $\sim$60~s, the neutrino light curve has a shallow decay phase. The $\rho_b$ profile does not evolve very much but the $s$ and $Y_e$ profiles vary in this phase. Meanwhile, the neutrinos are trapped inside the PNS and gradually leak out from the surface releasing the thermal energy. After about 60~s, since the neutrinoless $\beta$-equilibrium is achieved, the $Y_e$ profile becomes stationary and the neutrino luminosity reduces.

For the shallow decay phase, the neutrino light curve is sensitive to the EOS. In contrast, the dependence on the EOS is minor for early times of $t<10$~s. These features are consistent with the results of prior work by \citet{pons99}. Hereafter, we discuss the EOS dependence of the neutrino light curves for the shallow decay phase.

\begin{figure*}
\plotone{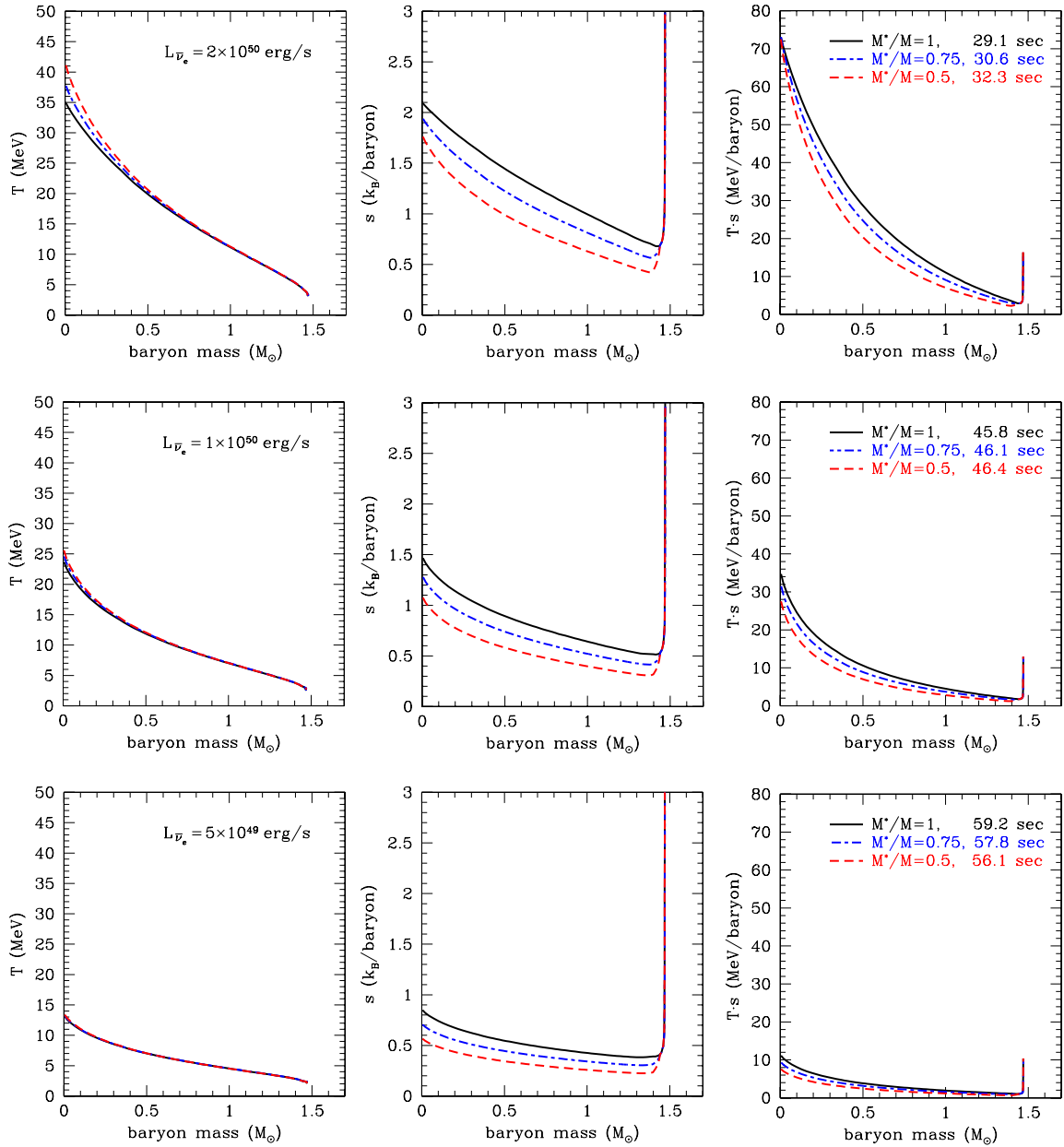}
\caption{Snapshots of the PNS cooling for our EOS models with $S_{00}=45$~MeV at the times when the luminosity of $\bar\nu_e$ is $2\times10^{50}$~erg~s$^{-1}$ (upper), $10^{50}$~erg~s$^{-1}$ (middle) and $5\times10^{49}$~erg~s$^{-1}$ (lower). The left, central and right panels show the profiles of temperature, entropy and the product of temperature and entropy, respectively. In all panels, solid (black), dot-dashed (blue) and dashed (red) lines correspond to the models with the effective masses of $u=1$, 0.75 and 0.5, respectively.}
\label{fig:udep}
\end{figure*}

\begin{figure*}
\plotone{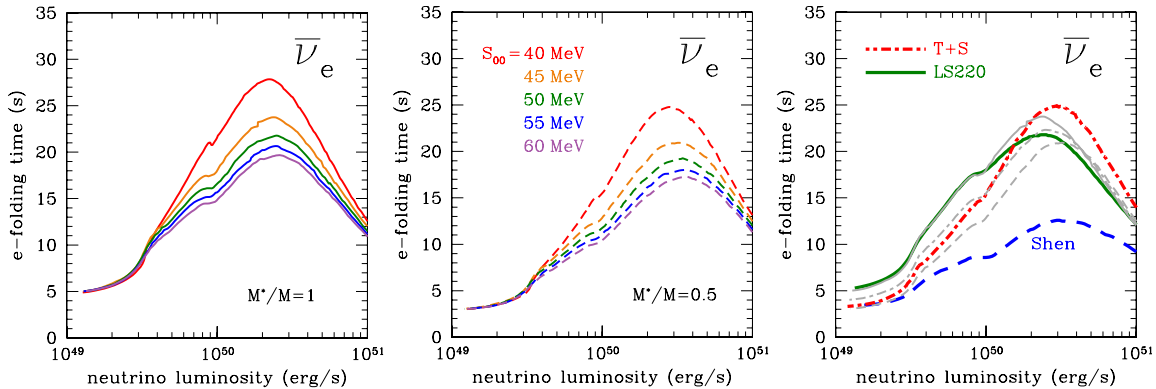}
\epsscale{1.0}
\caption{$e$-folding time of the $\bar \nu_e$ luminosity for our EOS models with effective masses of $u=1$ (left) and 0.5 (center), and for the LS220 EOS, Shen EOS and T+S EOS models (right). The plots are shown as a function of $\bar \nu_e$ luminosity. The notation of the lines is the same as that in Figure~\ref{fig:nulc}.}
\label{fig:efold}
\end{figure*}

From Figure~\ref{fig:nulc}, we can see that the cooling timescale is longer for the model with smaller $S_{00}$. Since the model with small $S_{00}$ has a soft EOS and a small PNS radius as stated in \S~\ref{sec:nseos}, the density is high as shown in Figure~\ref{fig:dens}, which shows the density profiles at the time when the $\bar \nu_e$ luminosity is $10^{50}$~erg~s$^{-1}$. Therefore, the neutrino mean free path is shorter and the cooling timescale is longer for the model with smaller $S_{00}$. Furthermore, the $S_{00}$ dependence of the neutrino light curves is stronger in the later phase because, in our EOS models, the difference in symmetry energy is large especially for high-density and low-$Y_e$ matter. In general, the uncertainties of the EOS increase with density and nuclear asymmetry while the nuclear properties are experimentally well established for the symmetric matter in the vicinity of the saturation density. In the process of supernova explosion, the matter has lower density and higher $Y_e$ for the first $\sim$1~s than cold neutron stars. Therefore, the EOS dependence is worth investigating for PNS cooling rather than for supernova explosion mechanism.

The symmetry energy of the EOS affects the neutronization of the PNS \citep[][]{sumi95}. The electron-type lepton number of the PNS is carried away by the net flux of electron-type neutrinos ($\nu_e-\bar\nu_e$). In Figure~\ref{fig:nulcd}, the difference between the number luminosities of $\nu_e$ and $\bar\nu_e$ is shown with the number luminosity of $\bar\nu_e$. The net flux of electron-type neutrinos is larger for the model with smaller $S_{00}$ because the electron fraction at the neutrinoless $\beta$-equilibrium is lower. Nevertheless, the difference between the number luminosities of $\nu_e$ and $\bar\nu_e$ is at most 10\% of their sum (right panel of Figure~\ref{fig:nulcd}). Therefore, in this study, the main effect of the symmetry energy on the neutrino emission from the PNS originates from the stiffness of the EOS. Incidentally, the symmetry energy of the EOS also affects the convective instabilities \citep[][]{robe12} while the convection is not taken into account in our simulation.

The cooling timescale also depends on the effective mass parameter $u$. In particular for the shallow decay phase, the model with large $u$ has a long cooling timescale. This is interpreted as follows. In the low-temperature regime, the entropy of baryons satisfies $s_b \propto uT$ and the thermal energy of baryons satisfies $E^{\rm th.}_b \propto uT^2$. Thus, the entropy and thermal energy are higher for the model with larger $u$ when the temperature is fixed. The neutrino luminosity depends on the temperature profile of the PNS. For the models with $S_{00}=45$~MeV, the $T$ and $s$ profiles are shown in Figure~\ref{fig:udep}. In this figure, we select the times when the $\bar \nu_e$ luminosity becomes $2\times10^{50}$~erg~s$^{-1}$, $10^{50}$~erg~s$^{-1}$ and $5\times10^{49}$~erg~s$^{-1}$ for each model. We can recognize that the $T$ profiles are similar to each other but the model with large $u$ has a high entropy. Since the evolutions of the $\rho_b$ profile and the PNS radius are minor in the shallow decay phase, the thermal energy divided by the neutrino luminosity roughly gives the cooling timescale. Furthermore, the thermal energy is estimated as $E^{\rm th.} \sim Ts$ (Figure~\ref{fig:udep}). Therefore, the cooling timescale is longer for a model with larger $u$ because the thermal energy stored in the PNS is larger.

Recently, the effective mass dependence has been studied by some authors in the context of supernova models \citep[e.g.,][]{schneider17,yasin18}. According to \citet{yasin18}, a more rapid contraction of the PNS is observed for the model with a larger effective mass. Although their result seems inconsistent with ours, this is not the case. Firstly, while they considered the evolution until about 1~s after the bounce, the cooling timescale investigated in our study is longer by two orders of magnitude. Secondly, in their EOS models, the effective mass is treated in a density-dependent manner and it is smaller at higher densities. As a result, their neutron star EOS is stiffer for the model with a smaller effective mass. In contrast, in our EOS model, the effective mass parameter $u$ is incorporated so as not to affect the neutron star EOS. Owing to this feature, the thermal contribution is separated from the variation of the high-density EOS. 

\begin{figure}
\plotone{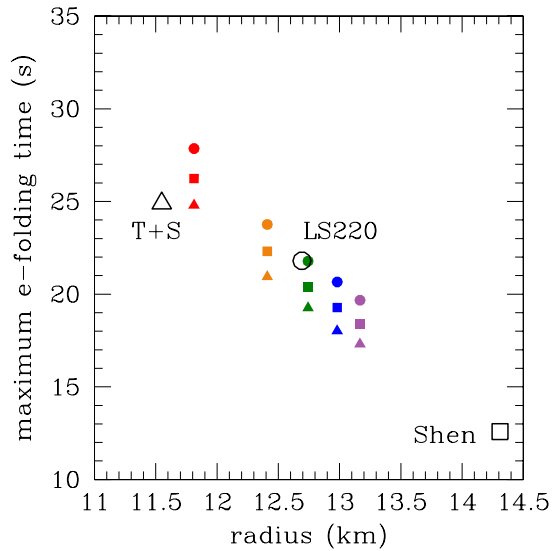}
\epsscale{1.0}
\caption{Relation between the radius of a cold neutron star with a baryon mass of $1.47M_\odot$ and the maximum $e$-folding time of $\bar \nu_e$ luminosity. The open circle, open square and open triangle represent the LS220 EOS, Shen EOS and T+S EOS, respectively. The filled circles, filled squares and filled triangles represent our EOS models with effective masses of $u=1$, 0.75 and 0.5, respectively.}
\label{fig:emax}
\end{figure}

So as to quantify the cooling timescale, we introduce the $e$-folding time of the $\bar \nu_e$ luminosity, $\tau_{\bar\nu_e}$, as
\begin{equation}
L_{\bar\nu_e}(t+\tau_{\bar\nu_e}) = \frac{L_{\bar\nu_e}(t)}{e},
\label{eq:efold}
\end{equation}
where $L_{\bar\nu_e}$ is the $\bar \nu_e$ luminosity and $e$ is the base of the natural logarithm. In Figure~\ref{fig:efold}, $\tau_{\bar\nu_e}$ is displayed as a function of $L_{\bar\nu_e}$. From this figure, we can see that $\tau_{\bar\nu_e}$ has a maximum value, which corresponds to the cooling timescale of the shallow decay phase. We plot the maximum value of $\tau_{\bar\nu_e}$ for each model as a function of the radius of the cold neutron star in Figure~\ref{fig:emax}, where the $S_{00}$ and $u$ dependences described above are confirmed. Note that the neutron star radius is an observable that indicates the stiffness of EOS. Thus, the neutron star with a smaller radius has a longer cooling timescale because the EOS is soft and the central density is high.

In Figure~\ref{fig:emax}, the relation between the neutron star radius and the cooling timescale is shown not only for our models but also for other supernova EOSs. The result of the LS220 EOS is in better agreement with the trend of our models with $u=1$ than those with $u<1$. This is consistent with the fact that the effective mass of nucleons is assumed to be equal to their rest mass in the LS220 EOS. In contrast, the Shen EOS and T+S EOS yield results consistent with $u<1$. Actually, the effective masses of these EOSs are lower than the rest mass and they depend on $\rho_b$, $T$ and $Y_e$. In particular, as $\rho_b$ increases and $Y_e$ decreases, the effective mass of neutrons decreases for the Togashi EOS \citep[][]{togashi13}, which is the high-density part of the T+S EOS. Therefore, the neutrino light curve of PNS cooling is expected to provide us an opportunity to probe the stiffness and effective mass of the nuclear EOS. For this purpose, evaluations of the neutrino event rate in neutrino detectors are required using not only various EOS models but also various PNS models with different masses and initial conditions \citep[][]{suwa19}.

\section{Conclusion}\label{sec:conc}

In this paper, we have systematically investigated the EOS dependence of PNS cooling for the first time. For this purpose, we have constructed a series of phenomenological EOSs, which have $S_{00}$ (the symmetry energy at a density of $2n_0$) and nucleon effective mass as control parameters. In the present model, the zero-temperature EOS and the radius of neutron stars depend only on $S_{00}$ and the effective mass parameter is incorporated so as not to affect the zero-temperature EOS. With $S_{00}=40$--60~MeV, our EOS models can account for the radius and tidal deformability of neutron stars indicated by GW170817. We have performed cooling simulations of PNSs with a baryon mass of $1.47M_\odot$, which corresponds to a gravitational mass of $\sim$1.33$M_\odot$. The numerical code utilized in this study follows quasi-static evolutions of PNSs solving the general-relativistic stellar structure with neutrino diffusion.

We have found that the cooling timescale is longer for the model with smaller $S_{00}$, that is, smaller symmetry energy at high densities. This is consistent with the fact that the PNS models with small $S_{00}$ have a short mean free path of neutrinos because the EOS is soft and the central density is high. We have also found that the cooling timescale is shorter for the model with smaller effective mass, for which the thermal energy stored in the PNS is smaller. So as to quantify the cooling timescale, we have introduced an $e$-folding time of the $\bar \nu_e$ luminosity and its maximum value. Then, we have compared the results obtained using our EOS models with those obtained using other supernova EOSs. It has been found that they are consistent in terms of their dependences on the effective mass and neutron star radius. Furthermore, provided that the effective mass is at least half of the rest mass and the neutron star radius is consistent with GW170817, the cooling timescale of our PNS models with a baryon mass of $1.47M_\odot$ has been predicted to be $\gtrsim$15~s.

The results in this paper imply that the future detection of supernova neutrinos will enable us to probe the properties of hot and dense matter inside a PNS whereas there are several issues that remain to be investigated. For this purpose, further systematic predictions of neutrino light curves are required for various PNS models with different masses and initial conditions. Improvements in the EOS and microphysics will also be interesting especially for the treatment of the temperature- and density-dependent effective mass. Since the neutrino interaction rates affect the quantitative results for the cooling timescale, the consistency of the neutrino opacities with the EOS is also an important future work. Nevertheless, we think that the qualitative trends found in this paper are correct. This paper will hopefully provide a basis for these forthcoming discussions.

\acknowledgments
The authors are grateful to Akira Ohnishi, Kazuhiro Oyamatsu, Kohsuke Sumiyoshi, Yudai Suwa, Hajime Togashi and Shoichi Yamada for valuable comments. In this work, numerical computations were partly performed on the supercomputers at Research Center for Nuclear Physics (RCNP) in Osaka University. This work was partially supported by JSPS KAKENHI Grant Numbers JP26104006 and JP17H05203.

\end{document}